\magnification=1000 \vsize=8.9truein \hsize=6.5truein \baselineskip=0.6truecm
\parindent=1truecm \nopagenumbers \font\scap=cmcsc10 \hfuzz=0.8truecm
\font\tenmsb=msbm10
\font\sevenmsb=msbm7
\font\fivemsb=msbm5
\newfam\msbfam
\textfont\msbfam=\tenmsb
\scriptfont\msbfam=\sevenmsb
\scriptscriptfont\msbfam=\fivemsb
\def\Bbb#1{{\fam\msbfam\relax#1}}

\null \bigskip  \centerline{\bf Linearisable systems and the Gambier approach }

\vskip 2truecm
\bigskip
\centerline{\scap S. Lafortune$^{\dag}$}
\centerline{\sl LPTM et GMPIB,  Universit\'e Paris VII}
\centerline{\sl Tour 24-14, 5$^e$\'etage}
\centerline{\sl 75251 Paris, France}
\footline{\sl $^{\dag}$ Permanent address: CRM, Universit\'e de
Montr\'eal, Montr\'eal, H3C 3J7 Canada}

\bigskip
\centerline{\scap B. Grammaticos}
\centerline{\sl GMPIB, Universit\'e Paris VII}
\centerline{\sl Tour 24-14, 5$^e$\'etage}
\centerline{\sl 75251 Paris, France}
\bigskip
\centerline{\scap A. Ramani}
\centerline{\sl CPT, Ecole Polytechnique}
\centerline{\sl CNRS, UPR 14}
\centerline{\sl 91128 Palaiseau, France}

Abstract
\medskip\noindent
A systematic study of the discrete second order projective
system is presented, complemented by the integrability analysis of the
associated multilinear mapping. Moreover, we show how we can obtain
third order integrable equations as the coupling of a Riccati  equation
with second order Painlev\'e equations. This is
done in both continuous and discrete cases.
\vfill\eject

\footline={\hfill\folio} \pageno=2

\bigskip
\noindent {\scap 1. Introduction}
\medskip
Integrability is far too general a term. In order to fix the ideas
we can just present three most common types of integrability, which suffice
in order to explain the properties of the majority of integrable systems
[1]. These three types are:

- Reduction to quadrature through the existence of the adequate number of
integrals of motion.

- Reduction to linear differential systems through a set of local
transformations.

- Integration through IST techniques. This last case is mediated  by the
existence of a Lax pair (a
linear system the compatibility of which is the nonlinear equation under
integration) which allows the
reduction of the nonlinear equation to a linear integrodifferential one.
The above notions can be extended {\sl mutadis mutandis} to the domain of
discrete systems.

This paper will focus on the second type of integrability, usually referred
to as linearizability. The
prototype of the linearizable equations is the Riccati. In differential
form this equation writes:
$$w'=\alpha w^2+\beta w+\gamma\eqno(1.1)$$
which is linearisable through a Cole-Hopf transformation.
Similarly, the discrete Riccati equation, which assumes the form of a
homographic mapping:
$$\overline x ={ {\alpha x+\beta}\over {\gamma x+\delta}}\eqno(1.2)$$
where $x$ stands for $x_n$, $\overline x$ for $x_{n+1}$ (and, of course,
$\underline x$ for $x_{n-1}$),
can be also linearized through a Cole-Hopf transformation.

The extension of the Riccati to higher orders can be and has been obtained
[2]. The simplest linearizable
system at $N$ dimensions is the projective Riccati which assumes the form:
$$w'_\mu=a_\mu+\sum_\nu b_{\mu\nu}w_\nu+w_\mu\sum_\nu c_\nu w_\nu\quad{\rm
with}\ \mu=1,\dots,N\eqno(1.3)$$
In two dimensions the projective Riccati system
can be cast into the second order equation:
$$w''= -3ww'-w^3+q(t)(w'+w^2)\eqno(1.4)$$
The discrete analog of the projective Riccati does exist and is studied in
detail in [3]. The
corresponding form is:
$$\overline x_\mu={a_\mu+x_\mu+\sum_\nu b_{\mu\nu}x_\nu\over 1-\sum_\nu
c_\nu x_\nu}\eqno(1.5)$$
Again in two dimensions the discrete projective
Riccati can be written as a
second-order mapping of the form
$$\alpha\overline x x  \underline x+\beta \overline x x+ \gamma x
\underline x +
\delta \overline x \underline x + \epsilon x + \zeta \overline x +\eta
\underline x
+\theta=0 \eqno(1.6)$$
which was first introduced in [4]. The coefficients
$\alpha,\beta,\ldots,\theta$ are not totally free.
Although the linearizability constraints have been obtained in [4], the
study of mappings of the
form (1.6) was not complete. In the present work we intend the study of
(1.6) in its general form.

 Another point must be brought to attention here.
In the continuous case the study of second order
equations has revealed the relation of the linearizable equations (1.4) to
the Gambier equation [5]. The
latter is obtained as a system of two coupled  Riccati in cascade
$$y'=-y^2+qy+c \eqno(1.7)$$
$$w'=aw^2+nyw+\sigma$$
where $n$ is an integer. It contains as a special case the linearizable
equation (1.4) which is obtained from (1.7) for $n=1$ and $a=-1$, $c=0$
and $\sigma=0$.
The discrete analog of the Gambier mapping was introduced in [6] and in
full generality in [7].

In the present work we shall also address the question
of the construction of integrable
third order systems in the spirit of Gambier. Namely we shall start with a
second order integrable equation and couple it with a Riccati (or a linear)
first order
(also integrable) equation. This enterprise may easily assume staggering
proportions. In order to limit the scope of our investigation we shall
consider coupled systems where the dependent variable enters only in a
polynomial way. This
leads naturally to the coupling of a Painlev\'e ($\Bbb P$) I or II to a
Riccati.

In the next section we shall analyse (1.6) and show how one can
isolate the integrable cases
through the use of the singularity confinement criterion. In Section 3
we will present how a third order integrable equation can be constructed
from the coupling of a Riccati and a Painlev\'e equation (in the discrete
and the continuous case).

\bigskip
\noindent {\scap 2. Linearizable mappings as discrete projective systems}
\medskip
In [4] we have introduced projective system as a way to linearize a
second-order mapping. (The general
theory of discrete projective systems has been recently presented in [3]).
In this older work of ours we
have focused on a three-point mapping that can be obtained from a
$3\times3$ projective system. The main idea was
to consider the system:
$$\pmatrix{\overline u\cr\overline v\cr \overline w}=
\pmatrix{p_{11}&p_{12}&p_{13}\cr p_{21}&p_{22}&p_{23}\cr p_{31}&p_{32}&p_{33}}
\pmatrix{u\cr v\cr w}\eqno(2.1)$$
and conversely
$$\pmatrix{\underline u\cr\underline v\cr \underline w}=
\pmatrix{m_{11}&m_{12}&m_{13}\cr m_{21}&m_{22}&m_{23}\cr m_{31}&m_{32}&m_{33}}
\pmatrix{u\cr v\cr w}\eqno(2.2)$$
where the matrix $M$ is obviously related to the matrix $P$ through
$\overline M=P^{-1}$ . Introducing
the variable
$x=u/v$ and the auxiliary $y=w/v$ we can rewrite (2.1) and (2.2) as
$$\overline x={p_{11}x+p_{12}+p_{13}y\over p_{21}x+p_{22}+p_{23}y}\eqno(2.3)$$
$$\underline x={m_{11}x+m_{12}+m_{13}y\over m_{21}x+m_{22}+m_{23}y}\eqno(2.4)$$
(Since the $m_{3i}$, $p_{3i}$ do not appear in (2.3), (2.4) and we can simplify
$M,P$ by taking $m_{33}=p_{33}=1$ and $m_{31}=p_{31}= m_{32}=p_{32}=0 $).
Finally eliminating $y$ between (2.3) and (2.4) we obtain the mapping:
$$\alpha\overline x x  \underline x+\beta \overline x x+ \gamma x
\underline x +
\delta \overline x \underline x + \epsilon x + \zeta \overline x +\eta
\underline x
+\theta=0 \eqno(2.5)$$
where the $\alpha, \beta, \ldots \theta $ are related to the $m, p$'s.

Equation (2.5) will be the starting point of the present study. Our
question will be when is an equation
of this form integrable? (Clearly the relation to the projective system
works only for a particular
choice of the parameters). In order to investigate the integrability of
(2.5) we shall use the
singularity confinement approach that was introduced in [8]. What are the
singularities of (2.5)? Given the form of (2.5) it is clear that diverging
$x$ does not lead to any
difficulty. However, another (subtler) difficulty arises whenever $x_{n+1}$
is defined independently of
$x_{n-1}$. In this case the mapping ``loses one degree of freedom''. Thus
the singularity condition is
$${\partial x_{n+1} \over \partial x_{n-1}}=0$$
which leads to :
$$(\alpha  x+\delta)(\epsilon x+\theta)=(\beta x+\zeta)(\gamma x+ \eta)
\eqno(2.6)$$
Equation (2.6) is the condition for the appearence of a singularity. Given
the invariance of (2.5) under
homographic transformations it is clear that one can use them in order to
simplify (2.6). Several choices
exist but the one we shall make here is to choose the roots of (2.6) so as
to be equal to $0$ and
$\infty$, unless of course (2.6) has two equal roots, in which case we
bring them both to $0$. Let us
examine the distinct root case. For the roots of (2.6) to be $0$ and
$\infty$ we must have:
$$\alpha \epsilon    =\beta\gamma \eqno(2.7)$$
$$\delta\theta=\zeta\eta$$
The generic mapping of the form (2.5) has $\alpha\theta\neq0$ and we can
take $\alpha=\theta=1$ (by
the appropriate scaling of $x$ and a division). We have thus,
$$\overline x x  \underline x+\beta \overline x x+ \gamma x \underline x
+\zeta\eta \overline x
\underline x + \beta\gamma x + \zeta \overline x +\eta \underline x
+1=0\eqno(2.8)$$
Nongeneric cases do exist as well and have been examined in detail in [9].

In order to investigate the integrability of the mapping (2.8) we shall
apply the singularity
confinement criterion.
Here the singularities are by construction 0 and $\infty$. Following the
results of [4] we require
confinement in just one step. We require to have an indeterminate form
$0/0$ at the step following the singularity.
This leads to the condition
$\beta=\zeta=0$. We thus obtain the mapping:
$$\overline x x \underline x + \gamma x \underline x+ \eta \underline x + 1
=0 \eqno(2.9)$$
or, solving for $\overline x$ :
$$\overline x = - \gamma + {\eta \underline x + 1\over x \underline x}
\eqno(2.10)$$
where $\gamma$ and $\eta$ are free.  This is indeed integrable: we can show
that (2.10) can be obtained from the
projective system (2.1), (2.2) provided we take
$$P=\pmatrix{-(\gamma q+1)&q \overline q&1\cr q&0&0\cr0&0&1}
\eqno(2.11)$$
and $M=\underline P^{-1}$ provided $q$ is some solution of the equation $
\overline q q
\underline q + q\underline q \eta + \underline q\, \underline \gamma + 1= 0 $.
Integrable but nonlinearizable cases of (2.8) do also exist: they have
been identified and presented in detail in [9].

Before closing this section let us present the  continuous limits of the
linearizable equation we
have identified above. For (2.9) we put $x = - 1 + \nu w$, $\gamma = 3+
\nu^2 p$, $\eta = \gamma +\nu^3 q$
and we obtain at the limit $\nu\to 0$ the equation:
$$w''= 3ww'-w^3+ p w + q \eqno(2.12)$$
This is equation \#6 in the Painlev\'e/Gambier classification [10] (in
noncanonical form) and precisely
the one that can be obtained from a $N=2$ projective Riccati system.

\bigskip
\noindent {\scap 3. Constructing Integrable Third Order Systems: the
Gambier Approach}
\medskip

The key idea of Gambier was to construct an integrable
second order equation by suitably coupling two integrable first order ones. The
latter were well-known: at first order the only integrable (in the sense of
having the
Painlev\'e property) ordinary differential equations are either linear or
of Riccati type. The
Gambier equation is precisely the coupling of two Riccati in cascade.

In [11], we extend this idea for third order systems. We couple Painlev\'e
second order equations with the Riccati
equation both in the continuous and the discrete cases.

The coupling of a Painlev\'e equation with a Riccati was first considered by
Chazy.  He examined the coupling P$_{\rm I}$:
$$w''=6w^2+z,$$
with a Riccati:
$$y'=\alpha y^2 +\lambda w+\gamma \eqno(3.1)$$
where $\alpha,\beta,\lambda,\gamma$ are functions of $z$.
This coupling is additive: it is indeed the
only coupling that is compatible with integrability.
Chazy found that (3.1) must have the form:
$$y'={1-k^2 \over 4}y^2 + w + \gamma, \eqno(3.2)$$
where $k=6m+n$.
Chazy found the following necessary integrability constraints:
$$
\matrix{
n=2 &\gamma    =0  \hfill\cr
n=3 &\gamma'  =0   \hfill\cr
n=4 &\gamma''  =\mu\gamma^2+\nu z  \hfill \cr
n=5 &\gamma''' =\mu\gamma\gamma' + \nu,\hfill \cr
}$$
where $\mu$ and $\nu$ are specific numerical constants. It turns out that
for $k=n$ they
are also sufficient. For
$k=6m+1$ the first condition appears at $k=7$. In this case the constraint
reads:
$$
\gamma^{(5)}=48 \gamma \gamma ''' +120 \gamma'\gamma''-{2304
\over 5}\gamma' \gamma^2 - 24z\gamma'-48\gamma.
$$
This equation has the Painlev\'e property and is thus expected to be
integrable.

In [11] we have presented the coupling of a Riccati to other integrable second
order differential equations.

To construct integrable discrete systems in the same spirit as Gambier we
need a detailed
knowledge of the forms of the equations to be coupled and an  integrability
detector. The second order mappings which play the role of the Painlev\'e
equations in the
discrete domain have been the object of numerous detailed studies and we
are now in possesion
of discrete forms of all the equations of the Painlev\'e/Gambier
classification. The discrete
integrability detector is based on the singularity confinement.

We consider the coupling of a discrete Riccati for the
variable $y$:
$$
\overline y = {(\alpha x + \beta)y +(\eta x + \theta) \over (\epsilon x +
\zeta)y +(\gamma x +
\delta)}, \eqno(3.3)
$$
(where  $\alpha$, $\beta$,
$\dots$,$\theta$ depend in general on $n$) the coefficients of which depend
linearly on $x$,
the solution of the discrete P$_{\rm II}$. The mapping (3.3) can be
brought under canonical form through the application of homographic
transformations on $y$ to either:
$$
\overline y = {(\alpha x + \beta)y +1 \over y +(\gamma x +
\delta)}. \eqno(3.4)
$$
or:
$$
\overline y (\gamma x +\delta)-y(\alpha x + \beta)-1=0. \eqno(3.5)
$$
In [11] we examined in detail the coupling of (3.4) and (3.5) with either
d-P$_{\rm I}$ or d-P$_{\rm II}$ (under various forms).
Here, we will focus on a particular example of a coupling to
d-P$_{\rm II}$.

How does one apply the singularity confinement criterion to a mapping such
as (3.4) when $x$
is given by some discrete equation like d-P$_{\rm I}$ or d-P$_{\rm II}$? The
singularity manifests itself by the fact that $\overline y$ is independent
of $y$ i.e. when
$$
(\gamma x + \delta)(\alpha x + \beta)=1. \eqno(3.6)
$$
This quadratic equation has two roots $X_1$, $X_2$.  The confinement
condition is for $y$
to recover the lost degree of freedom. This can be done if $y$ assumes an
indeterminate form
$0/0$. This means that $x$ at this stage must again satisfy (3.6) and
moreover be
such that the denominator (or, equivalently, the numerator) vanishes.

Let us assume now that for some $n$ we have $x_n=X_1$.
 The confinement requirement is that $k$
steps later $x_{n+k}=X_2$. Starting from $x_n=X_1$ and some initial datum
$x_{n-1}$, we can iterate the mapping for $x$ and obtain $x_{n+k}$ as a
complicated function
of $x_{n-1}$ and $X_1$. Since
$x_{n+k}$ depends on the free parameter $x_{n-1}$ there is no hope for
$x_{n+k}$ to be
equal to $X_2$ if $X_1$ is a generic point for the mapping of $x$.
The only possibility is
that both
$X_1$ and
$X_2$ be special values. What are the special values of this equation
depends on its details, but clearly in the case of the discrete Painlev\'e's
we shall examine here, these values can only be the ones  related to the
singularities. To be more
specific, let us examine d-P$_{\rm II}$:
$$
\overline x + \underline x = {zx+a \over 1-x^2}. \eqno(3.7)
$$
The only special values of $x$ are the ones related to the singularity
$x_n=\pm 1$,
$x_{n+1}=\infty$, $x_{n+2}=\mp 1$ while $\dots$, $x_{n-2}$, $x_{n-1}$ and
$x_{n+3}$, $x_{n+4}$,
$\dots$ are finite. This means that the two roots of (3.6) must be two of
$\{+1, \infty,
-1\}$  and moreover that confinement must occur in two steps. The precise
implementation of singularity confinement requires that the denominator of
(3.4) at $n+2$
vanishes (and because of (3.6) this ensures that the numerator vanishes as
well). Moreover,
we must make sure that the lost degree of freedom (i.e. the dependence on
$y$) is indeed
recovered through the indeterminate form.

The singularity patterns of (3.7) are  $$\{\pm
1,\infty,\mp 1\}. \eqno(3.8)$$
This means that the singularity condition (3.4) must have $\pm 1$ as roots.
As a result we
have:
$$\delta=-\beta/(\alpha^2-\beta^2), \eqno(3.9)$$
$$\gamma=\alpha/(\alpha^2-\beta^2). \eqno(3.10)$$
The two different patterns lead to a first confinement
conditions given by:
$$\beta=k\alpha$$
 where $k$ is a constant with binary freedom which we
will ignore from now. The second condition:
$$
\underline \alpha \alpha^2 \overline \alpha = {1 \over (1-k^2)^2}. \eqno(3.11)
$$
This equation can be solved by linearisation just by taking the logarithm
of both sides. More examples of couplings of discrete equations can be found
in [11].

\bigskip
\noindent {\scap 3. Conclusion}
\medskip
\noindent
In the previous sections we have first investigated 3-point mappings that are
integrable
through linearization. Our analysis was guided by the analogy with the
continuous
situation and results of ours on $N=2$ projective systems. We
have presented an analysis of the linearizable mapping and
identified one of its
integrable form.  Moreover we have presented an approach for the
construction of integrable
 third order systems through the coupling of a second order equation to a
Riccati or a
linear first order equation. Thus we have extended the Gambier approach
(first used in his
derivation of the second order ODE that bears his name) to higher order
systems. We have
applied this coupling method to both continuous and discrete systems.
\bigskip

\bigskip
{\scap References}
\smallskip
\item{[1]} M.D. Kruskal, A. Ramani and B. Grammaticos, 
{\it Singularity analysis
and its relation to complete, partial and non-integrability}, 
NATO ASI Series C 310, Kluwer 1989, 321-372.
\item{[2]} R.L. Anderson, J. Harnad and P. Winternitz, {\it Systems of 
ordinary differential equations with superposition principles}, 
Physica D4 (1982) 164-182.
\item{[3]} B. Grammaticos, A. Ramani and P. Winternitz, {\it 
Discretizing families of linearizable equations},
 Phys.Lett.A 245 (1997), 382-388.
\item{[4]} A. Ramani, B. Grammaticos and G. Karra, 
{\it Linearizable mappings}, Physica A 181 (1992) 115-127.
\item{[5]} B. Gambier, {\it 
Sur les \'equations diff\'erentielles du second ordre et du premier 
degr\'e dont
l'int\'egrale g\'en\'erale est \`a points critiques fixes}, 
Acta Math. 33 (1910) 1-55.
\item{[6]} B. Grammaticos and A. Ramani, {\it The Gambier mapping}, 
Physica A 223 (1996) 125-136.
\item{[7]} B. Grammaticos, A. Ramani and S. Lafortune, {\it The Gambier
mapping, revisited}, Physica A 253 (1998) 260-270.
\item{[8]} B. Grammaticos, A. Ramani and V. Papageorgiou, {\it Do
integrable mappings have the Painlev\'e property?}, Phys. Rev. Lett.
67 (1991), 1825-1828.
\item{[9]} B.Grammaticos, A.Ramani, K.M.Tamizhmani and S.Lafortune, 
{\it Again, linearisable mappings}, Physica A 252 (1998), 138-150.
\item{[10]} E.L. Ince, {\it Ordinary differential equations}, Dover, New
York, 1956.
\item{[11]} S. Lafortune, B. Grammaticos and A. Ramani, {\it Constructing 
third order integrable systems: the Gambier approach}, Inverse Problems 
14 (1998), 287-298.
\end